\documentclass[a4paper]{article}

\usepackage{INTERSPEECH2020}
\usepackage{algorithm}
\usepackage{subfigure}
\usepackage{multirow}
\usepackage{multicol}
\usepackage[noend]{algpseudocode}

\title{Low Latency End-to-End Streaming Speech Recognition with a Scout Network }
\name{	{Chengyi Wang$^{\diamondsuit}$, Yu Wu$^{\diamondsuit}$, Liang Lu$^{\dagger}$, Shujie Liu$^{\diamondsuit}$, Jinyu Li$^{\dagger}$,   Guoli Ye$^{\dagger}$, Ming Zhou$^{\diamondsuit}$}}
\address{$^{\diamondsuit}$ Microsoft Research Asia, Beijing \\
	$^{\dagger}$Microsoft Speech and Language Group }
\email{\{v-chengw, Wu.Yu, shujliu, jinyli,  lial, guoye, mingzhou\}@microsoft.com}

\begin{document}
\maketitle
\begin{abstract}
The attention-based Transformer model has achieved promising results for speech recognition (SR) in the offline mode. However, in the streaming mode, the Transformer model usually incurs significant latency to maintain its recognition accuracy when applying a fixed-length look-ahead window in each encoder layer. In this paper, we propose a novel low-latency streaming approach for Transformer models, which consists of a scout network and a recognition network. The scout network detects the whole word boundary without seeing any future frames, while the recognition network predicts the next subword by utilizing the information from all the frames before the predicted boundary. Our model achieves the best performance (2.7/6.4 WER)  with only 639 ms latency on the test-clean and test-other data sets of Librispeech.



\end{abstract}

\section{Introduction}
\begin{figure*}[t]
    \centering
    \includegraphics[width=0.9\textwidth]{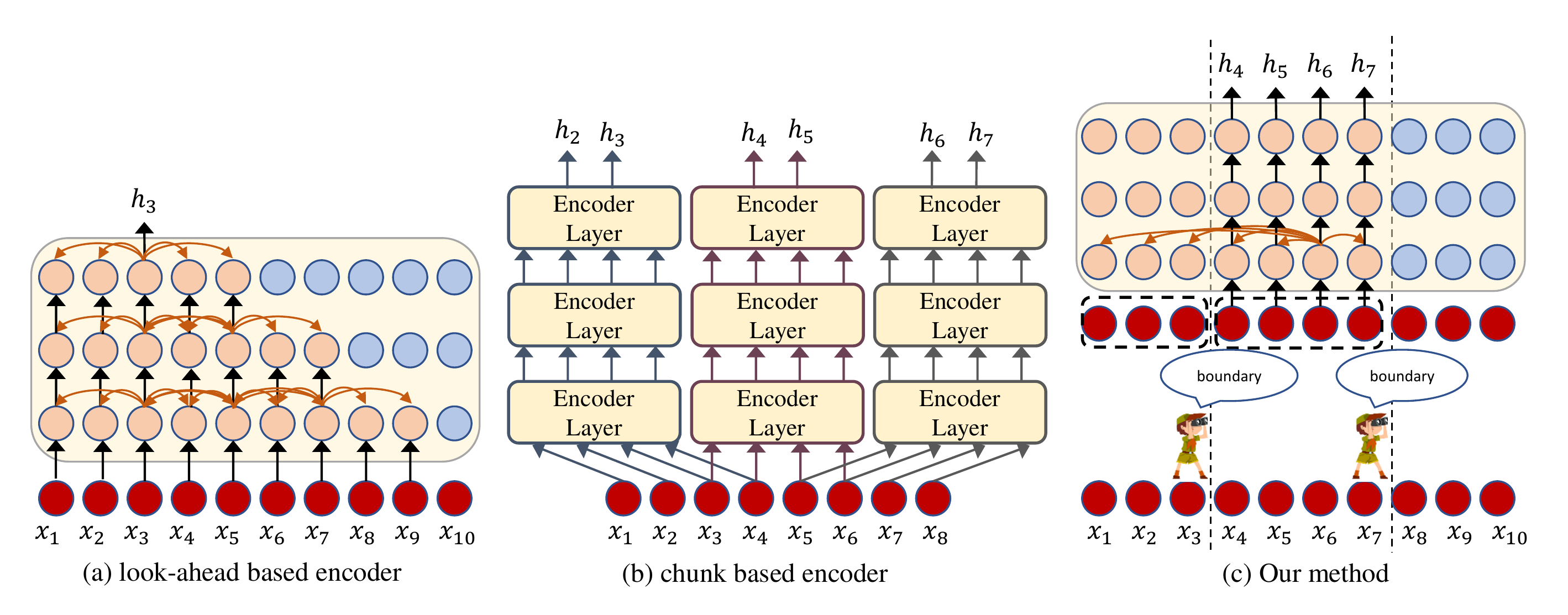}
    \caption{A comparison between three Transformer based streaming models.}
    \label{fig:main_pic}
    \vskip-4mm
\end{figure*}

Recently, there has been a surge of end-to-end (E2E) automatic speech recognition (ASR) models, including the connectionist temporal classification (CTC)  \cite{graves2006connectionist,graves2014towards,amodei2016deep}, the RNN-Transducer  \cite{graves2012sequence,graves2013speech,rao2017exploring} and the attention-based encoder-decoder (AED) models  \cite{chorowski2015attention,chan2016listen,chiu2018state,watanabe2017hybrid} due to their simple training procedure, desirable decoding efficiency and promising performance on large-scale speech benchmarks. In particular, the Transformer model  \cite{DBLP:conf/nips/VaswaniSPUJGKP17} has been successfully applied to the E2E ASR, which enjoys faster training and better performance compared with RNNs.  However, it is nontrivial to apply Transformer in an online recognition system due to the global encoder-decoder attention mechanism and the self-attention encoder, which requires the entire utterance. To solve the first problem, Transformer based monotonic chunkwise attention (MoChA)  \cite{DBLP:journals/corr/abs-1910-11871} and trigger attention mechanism (TA)  \cite{DBLP:journals/corr/abs-2001-02674} have been proposed to replace the global encoder-decoder attention. Regarding the streaming encoder, existing approaches can be categorized into look-ahead based method  \cite{DBLP:journals/corr/abs-2001-02674,DBLP:journals/corr/abs-2002-02562} and chunk-based method  \cite{DBLP:conf/icassp/DongWX19, DBLP:journals/corr/abs-1910-11871, DBLP:journals/corr/abs-1912-02958, DBLP:journals/corr/abs-2001-08290}, as shown in Figure \ref{fig:main_pic}(a) and Figure \ref{fig:main_pic}(b). The former sets a look-ahead window for each frame to incorporate the necessary context information. However, latency will increase linearly with the number of layers. The chunk-wise approach segments the utterance into several fixed-length chunks and feeds them to the encoder one by one. There are always overlaps between chunks to improve the performance. However, this hurts the training parallelism of self-attention layers, and degrades the recognition accuracy when the chunk size is small.

In this paper, we propose an adaptive look-ahead approach to trade-off latency and WER, where the context window size is modified dynamically. We hypothesize that the most valuable contextual information to produce an output token is from the speech segment which corresponds to this word.  Therefore, we introduce a neural component to detect the boundaries in the speech where a word starts and stops. We refer to this component as the scout network (SN), with a metaphor as a scout sent out ahead of the main force to gather the valuable information. Then the recognition network (RN) only looks ahead the frames up to the detected word boundaries. In this way, the recognition latency is not fixed but dependent on the duration of words and the segmentation performance of the SN. 

To train the SN, we formulate it as a simple sequence labeling problem, where each frame is classified as either a boundary or not. We use the force-alignment results as ground truth to train the network. SN does not see any future information for the boundary detection. Hence, there is no additional latency overhead in the SN. After the word boundaries are detected, any end-to-end (E2E) model can be employed as the RN. In this paper, we use the TA based Transformer model  \cite{DBLP:journals/corr/abs-2001-02674} as the RN to conduct frame-synchronous one-pass decoding by looking ahead to the detected word boundaries. 

Our experiments are conducted on Librispeech benchmark  \cite{DBLP:conf/icassp/PanayotovCPK15}. The results show that our proposed method can not only significantly reduce the latency, but also achieve the state-of-the-art recognition quality. Our base model with 78M parameters and large model  with 138M parameters achieve 2.9/7.4 and 2.7/6.4 on test-clean and test-other datasets.
To understand the effect of our proposed SN, we conduct experiments to analyze the relationship between the word error rate of the RN and the threshold in the SN. Experiments show that the higher threshold of the prediction precision ( with lower recall of the boundaries, and higher latency), leads to better recognition quality. The latency and the quality of the recognition can be balanced with the precision threshold of the SN.

\section{Background}
\subsection{Transformer ASR}
Transformer achieves promising results in ASR  \cite{DBLP:journals/espnet,wang2019transformer}.  Given a $T$-length speech feature sequence $\mathbf{X}$, the encoder transforms it to an  intermediate representation $\mathbf{H}$, then the decoder predicts the following word $y_i$ based on $\mathbf{H}$ and previous outputs $\mathbf{Y}_{[1:i-1]}$.

The Transformer encoder consists of a convolution block and $N_e$ encoder blocks, each of which has a multihead self-attention layer and a feedforward layer. The decoder is composed of $N_d$ decoder blocks, including a self-attention layer, an encoder-decoder attention layer and a feed-forward layer. Multihead attention mechanism is proposed, where in each head, weights are formed from queries ($\mathbf{Q} \in \mathcal{R}^{d}$) and keys ($\mathbf{K} \in \mathcal{R}^d$) and then applied to values ($\mathbf{V} \in \mathcal{R}^d$) as
\begin{small}
\begin{flalign}
\text{Multihead}(\mathbf{Q,K,V}) = \text{Concat}(\rm{head}_{1:m})\mathbf{W}^{head} \\
\text{~where~}\rm{head_i} = \text{softmax}(\frac{\mathbf{QW_qW_k^{T}K^{T}}}{\sqrt{d}})\mathbf{VW_v}
\end{flalign}
\end{small}
$\mathbf{Q}$, $\mathbf{K}$ and $\mathbf{V}$ have the same dimension $d$. $m$ is the number of attention heads.
Residual connections and layer normalization are applied for each block.

\subsection{Streaming ASR}
To build a streaming AED based ASR system, both the encoder and the decoder are allowed to access limited future context. For the Transformer encoder, the self-attention mechanism provides a flexible way to control the range of context by masking the attention scores \cite{DBLP:journals/corr/abs-2001-02674,DBLP:journals/corr/abs-2002-02562}. However, the receptive field and latency will increase linearly with the number of layers. As shown in Figure \ref{fig:main_pic}(a), suppose the right look-ahead window size is $w_r$, the latency is $N_e \times w_r$ frames.
Motivated by Transformer-XL  \cite{DBLP:conf/acl/DaiYYCLS19}, other works  \cite{DBLP:journals/corr/abs-1910-11871,DBLP:conf/icassp/DongWX19,DBLP:journals/corr/abs-1912-02958,DBLP:journals/corr/abs-2001-08290,DBLP:conf/asru/TsunooKKW19} use a chunk-wise approach, where the entire utterance is segmented into several fixed-length chunks as shown in Figure \ref{fig:main_pic}(b). 

In terms of the decoder, the key is to learn the online monotonic alignments.  \cite{DBLP:conf/icassp/BahdanauCSBB16,DBLP:conf/interspeech/ChanL16} restricts the range of the attention to a fixed-size window with position determined by previous attention distribution. Monotonic chunkwise attention (MoChA)  \cite{DBLP:conf/iclr/ChiuR18} and its extension  \cite{DBLP:journals/corr/abs-2001-08290,fan2018online} use a trainable energy function to shift the window.  \cite{DBLP:conf/asru/TsunooKKW19} applies it to multi-head encoder-decoder attention.  \cite{DBLP:conf/icassp/MoritzHR19} proposes trigger attention where a CTC module triggers the computation of the attention. 


In this work, we use a Scout Network to detect the boundaries of words, enabling the encoder and the decoder to have an adaptive right context window. We believe the context window to the word boundary provides the most useful information for recognition and brings a better tradeoff between latency and performance. 

\section{Method}

\subsection{Scout Network}
The purpose of the SN is to perform word boundary detection. In principle, SN can be implemented using any type of neural networks, while in this work, we use a similar architecture as the ASR encoder. It uses CNN as the pre-processing layer with a sampling rate $r$, followed by $N_s$ self-attention layers. Given the  acoustic input sequence $\mathbf{X}=(\bm {x}_1, \dots, \bm {x}_T)$, in each self-attention layer, the output of the current frame $i$ is conditioned only on the previous states, so that there is no additional latency incurred in the SN. we denote the output hidden sequence as $\mathbf{H}^S=(\bm {h}^s_1, \dots, \bm {h}^s_{T'})$, where $T' = \lceil \frac{T}{r} \rceil$. Then a linear layer and a sigmoid layer predicts a boundary probability $p_i$.
We train the SN by minimizing the cross-entropy loss as:
\begin{small}
\begin{equation}
    \mathcal{L}_{SN} = \sum_{T'}{b}_i\text{log}(p_i) = \sum_{T'}{b}_i\text{log}(\text{Sigmoid}(\mathbf{W}\bm{h}_i^s))
    \vspace{-2mm}
\end{equation}
\end{small}
where $b_i \in \{0, 1\}$ is the ground truth of the boundary decision. In this work, we used the Montreal Forced Aligner  \cite{mcauliffe2017montreal} to perform word-level force-alignment and obtained the labels. During inference of the SN, we set a threshold $\sigma \in (0, 1)$ to determine whether a frame is a boundary or not. Once $p_i \geq \sigma$, we denote the $i$-th frame as a boundary, that is $b_i=1$. Otherwise, $b_i$ is set to 0. $\sigma$ is tuned on the validation set. 




\subsection{Streaming ASR with the Scout Network}
\subsubsection{Recognition Network Training}



For the recognition network encoder, the CNN pre-processing block is the same as in the SN. Thus, RN and SN have the same downsampling rate. During training of RN, we sample $\tilde{b}$ according to the probability $p$ produced by the SN as the predicted boundary. Suppose $\tilde{b}_{g_j}=1$ is the $j$-th predicted boundary, we note $g_j$ as the end time tag of the $j$-th word. Then we update the $i$-th hidden state in  $l$-th self-attention layer as follows: 
\begin{flalign}
\vspace{-2mm}
    \bm \tilde{h}^l_i = \text{Multihead}(h^{l-1}_i, \mathbf{H}^{l-1}_{[1:g_j]}, \mathbf{H}^{l-1}_{[1:g_j]})
\end{flalign}
where $i \in (g_{j-1}, g_j)$. The $i$-th state can only access the states up to the time step $g_j$. $\mathbf{H}^{l-1}_{[1:g_j]}$ is regarded as key and value. It can also be seen as chunk-based encoder where the chunk size is adaptive and there is no overlap between chunks.  During training, we rely on the mask strategy to avoid the model seeing context after the $g_j$-th frame.

We can use any type of the streaming decoder in our recognition network such as MoChA or Triggered Attention (TA)~ \cite{DBLP:conf/icassp/MoritzHR19}. In this work, we used TA with an adaptive look-ahead window in the decoder.  Once the CTC predicts a token $y_k$ at frame $i$, the decoder is triggered and the encoder-decoder attention is computed based on the segment of encoder output states $\mathbf{H}_{[1:g_j]}$. It should be noted that $j$ may not equal to $k$ due to prediction error of SN and the text tokenization. We weighted average the CTC loss and the seq2seq loss to train our model:
\begin{align}
    \mathcal{L} = - \gamma \log P_{s2s}(\mathbf{Y}|\mathbf{X}) - (1-\gamma) \log P_{ctc}(\mathbf{Y}|\mathbf{X}). \\
    \text{where~} {P_{s2s}(\mathbf{Y}|\mathbf{X})} = \sum_{y_k \in \mathbf{Y}}P_{s2s}(y_j|\mathbf{Y}_{[1:j-1]}, \mathbf{H}_{[1:g_j]})
\end{align} We set $\gamma$ as 0.7 in this work. 
The model is initialized from an offline Transformer and fine-tuned in a streaming manner.

To train the decoder, the alignment between CTC paths and the label sequence $\mathbf{Y}$ is required. Different from  \cite{DBLP:journals/corr/abs-2001-02674} which performs Viterbi alignment during training, we used the path with the highest Viterbi alignment score generated by the pre-trained offline model to trigger the decoder.

\subsubsection{Decoding}
\begin{algorithm}[th] \small
\caption{Streaming Transformer Decoding with Scout Network}
\label{alg:decode}
\begin{algorithmic}[1]
\Procedure{Scout-then-Decode}{$\mathbf{X}$, $ K, \sigma, \sigma_0, \lambda, \alpha, \beta$ }
\State $k \gets 0$, $g_k \gets 0$  
\State $\ell \gets (\langle sos \rangle,)$, $\Omega \gets \{l\}$

\For {i = 0 to $T$}
    \State $p_{i'} \gets \text{SN}(x_i$), $i' \gets \frac{i}{r}$ \Comment{Scout Boundary}
    \If{$p_{i'} > \sigma$}
        \State $k \gets k + 1$, $g_k \gets i'$
        \State $\mathbf{H}_{[g_{k-1}+1:g_k]}\gets \textbf{ENC}(\mathbf{X}_{[1: i]}$)
        \State $\Omega, p_{\text{joint}} \gets $\textsc{Decode}($\Omega$, $\mathbf{H}_{[1:g_k]}$, $g_{k-1}+1$, $g_k$,  $K$, $\sigma_0$,  $\lambda$, $\alpha$, $\beta$)
    \EndIf
    \EndFor
\State \textbf{return} $\textsc{Max}(\Omega, p_{\text{joint}}, 1)$
\EndProcedure

\Procedure{Decode}{$\Omega$, $\mathbf{H}$, start, end, $K$, $\sigma_0$, $\lambda$, $\alpha$, $\beta$}
\State $\hat{\Omega}_{\text{ta}} \gets \varnothing$
\For {j = start to end}
\State $\Omega_{\text{ctc}}, p_{\text{ctc}} \gets \textsc{CTCPrefix}(\Omega, \sigma_0, \bm {h}_j)$
\For {$\ell$ in $\Omega_{\text{ctc}}$}
    \If{$\ell$ not in $\hat{\Omega}_{\text{ta}}$}
        \State $p_{\text{ta}}(\ell) \gets P_{s2s}(\ell|\mathbf{H})$
        \State \text{add $\ell$ to $\hat{\Omega}_{\text{ta}}$}
    \EndIf
     \State $p_{\text{local}}(\ell) \gets \text{log~} p_{\text{ctc}} + \alpha \text{log~} p_{\text{LM}} + \beta |\ell|$
    \State $p_{\text{joint}}(\ell) \gets \lambda \text{log~} p_{\text{ctc}} + (1-\lambda) \text{log~}p_{\text{ta}} + \alpha \text{log~} p_{\text{LM}} + \beta |\ell|$

\EndFor

\State $\Omega_{\text{local}} \gets \textsc{Max}(\Omega_{\text{ctc}}, p_{\text{local}}, K)$ \Comment{Beam Pruning}
\State $\Omega_{\text{ta}} \gets \textsc{Max}(\Omega_{\text{ctc}}, p_{\text{ta}}, K)$
\State $\Omega_{\text{joint}} \gets \textsc{Max}(\Omega_{\text{ctc}}, p_{\text{joint}}, K)$
\State $\Omega \gets \Omega_{\text{local}} \cup \Omega_{\text{ta}}\cup\Omega_{\text{joint}}$

  \EndFor
\State \textbf{return} $\Omega$, $p_{\text{joint}}$
\EndProcedure
\end{algorithmic}
\end{algorithm}

Algorithm \ref{alg:decode} gives the decoding procedure for streaming Transformer with the Scout Network. The hyper-parameters used by the function include beam width $K$, boundary decision threshold $\sigma$, CTC decoding threshold $\sigma_0$, CTC decoding weight $\lambda$, language model weight $\alpha$ and length penalty $\beta$. In the 2nd line of the algorithm, we first initialize the index of speech segment $k=0$ and the end boundary of the $k$-th segment $g_k$.  The hypothesis set $\Omega$ is initialized in line 3 with the prefix sequence $\ell$ containing only the start of the sequence label $\langle\text{sos}\rangle$. In line 4-6, the frame-level feature is fed into the Scout Network and an instantaneous decision is made by threshold $\sigma$. Once a word boundary is detected, the ASR decoding is triggered at line 8-9.  

The \textsc{Decode} function shown from line 11 to 25 is similar to the decoding scheme in  \cite{DBLP:journals/corr/abs-2001-02674} with a small modification in beam pruning. In line 14, we perform the CTC prefix beam search  \cite{hannun2014first} based on the current encoder state $\bm h_j$ and prefix set $\Omega$, generating candidates set $\Omega_{\text{ctc}}$. Then the trigger attention decoder scores every candidate. To avoid duplicated computation, we store all the scored candidates in $\hat\Omega_{\text{ta}}$. In line 19 to 20, a local score and a joint score are assigned to each candidate. Then we select the top-$K$ candidates based on local score, attention score and joint score respectively, and combine them as the hypothesis set for the next step. When the process finishes, we select the prefix with the highest joint score as the decoding output.

\section{Experiments}
\subsection{Setup}

We performed our experiments on the LibriSpeech dataset  \cite{DBLP:conf/icassp/PanayotovCPK15}, which contains 960 hours of audio in the training set. We use {{dev-clean}} as the validation set and report results on the {{test-clean}} and {{test-other}} sets. 

Our approach is implemented based on ESPnet  \cite{DBLP:journals/espnet}. We used 80-dim log Mel-filter bank features with 3-dim pitch features \cite{DBLP:conf/icassp/GhahremaniBPRTK14} with  10ms sampling rate. The text is tokenized using SentencePiece \cite{kudo2018subword} and we set the vocabulary size to 5000. We evaluated our method in two settings, the Base setting and the Large setting. For the Base setting, we used the same architecture as  \cite{DBLP:journals/corr/abs-2001-02674} with $d_{model}=512, d_{ff}=2048, d_h=4, N_e=12$, and $N_d=6$ for a fair comparison to previous works. We used the released Transformer model provided by ESPnet\footnote{{https://github.com/espnet/espnet/tree/master/egs/librispeech/asr1}} as the pretrained offline model. For the Large setting, we used the architecture described in our previous work  \cite{wang2019semantic} with $N_e=24$ and $N_d=12$. 
For both settings, the down-sampling rate $r$ is 4, so that the self-attention layers operate at the sampling rate of 40ms per frame. We updated the SN using Adam optimizer with a learning rate of 0.001. The RN was trained using a warmup step of 2500 and a learning rate coefficient of 1.0 following the schedule in  \cite{DBLP:conf/nips/VaswaniSPUJGKP17}. SpecAugment \cite{park2019specaugment} is applied following the recipe in ESPnet. We run the fine-tuning stage for about 20 epochs. For decoding, we averaged the last 5 checkpoints as the final model. For decoding hyper-parameters, we set $K=10, \sigma_0=0.0005, \lambda=0.5, \alpha=0.5, \beta=2.0$ as in Algorithm \ref{alg:decode}. The N-best was re-scored using an LSTM language model provided by ESPnet.

\subsection{Scout Network Evaluation}
\begin{figure}[t]
    \centering
    \includegraphics[height=0.35\textwidth]{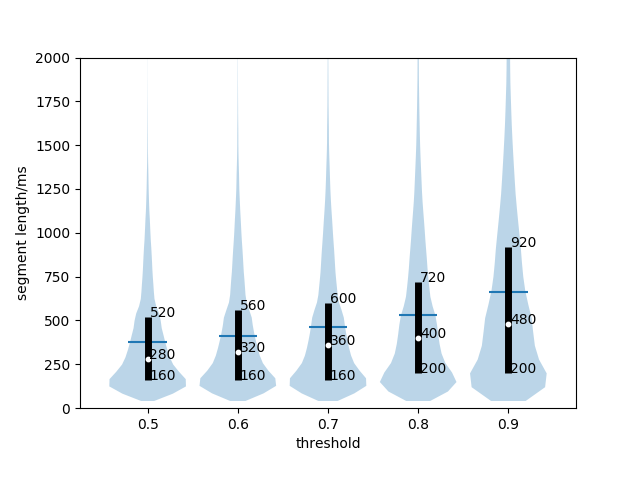}
    \vskip-4mm
    \caption{The distribution of segment length for different threshold. The blue area shows the density. The deep blue lines represent the mean values. The white markers represent the median values and the black vertical lines represent the interquartile ranges.  }
    \label{fig:distribution}
    \vskip-4mm
\end{figure}

\begin{table}
    \centering
     \caption{An example of the accuracy evaluation for the Scout Network. $E'$ and $E$ are downsampled positions.}
    \begin{tabular}{l|l}
    \hline
        $Y$ & it gave an imposing appearance to  \\
             & most of the wholesale houses  \\ \hline
         $E'$ &  7~ 12~ 14~ 28~ 42~ 45~ 52~ 54~ 56~ **~ 68~  \\ \hline
         $E$ &  7~ 12~ 15~ 28~ 42~ 45~ 52~ **~ 56~ 61~ 68~ \\ \hline
         evaluate & ~ ~ ~ ~ \, sub  ~ ~ ~ ~ ~ ~ ~ ~ ~ ~ ~ ~ del ~ ~ ~ \,ins \\ \hline
         word latency (ms) & 30 30 70~ 30~ 30~ 30~ 30~ 110 30~ --~ \, 30              \\ \hline
    \end{tabular}
     \label{tab:example}
     \vskip-4mm
\end{table}
We first evaluate the accuracy of the Scout Network for word boundary prediction. Given a predicted boundary $E$ and a reference boundary $E'$ from the force-alignment, we use the edit distance as the metric, which can be decomposed into  substitution (sub), deletion (del), insertion (ins) rates. An example is given in Table \ref{tab:example}. In Table \ref{tab:segmentation evaluation}, we show the validation results of different number of layers $N_s$ and threshold $\sigma$. We observe a high precision but low recall with a high threshold such as $\sigma = 0.9$. And not surprisingly, a shallower Transformer degrades the accuracy of the SN slightly. To draw the connection between the threshold and the latency of the recognition network, we further evaluate the distribution of the segment lengths on the joint test sets, which is shown in Figure \ref{fig:distribution}, where we used the base SN with $N_s=12$. For each segment, the length is computed as $(g_j - g_{j-1}) \times 40$ms, where $g_j$ is the predicted end boundariy. As shown in the figure, even with a threshold of 0.9, there are still 50\% segments shorter than 480ms and 75\% segments shorter than 920ms. Though some segments have a length longer than $2$s, we can control the maximum length through rules in practice. 
We also list the scout network efficiency running on a P40 GPU and 24 cores  Intel(R) 2.60GHz CPU in Table \ref{tab:segmentation evaluation}. The computation time is less than the duration of each frame (40ms), showing that the SN  is not a burden in decoding.

\begin{table}[t]
  \centering
    \caption{Segmentation Evaluation on the dev clean dataset. The first two columns are averaged per frame time cost.}
  \begin{tabular}{l|cc|c|ccc}
    \hline
    $N_s$  & GPU & CPU      & $\sigma$     & sub  &  del &  ins  \\ \hline
  &   & & 0.5   & 13.0  & 7.2  &1.9    \\
     12   &  11.1ms    & 29.8ms  &  0.7  & 8.0  & 21.0  & 0.6           \\
    &    &     &  0.9  &  2.6  & 42.6   & 0.1           \\  \hline
    8  & 7.4ms  & 19.7ms & 0.9  &  2.5  & 51.4  & 0.1 \\ \hline
    4  & 3.7ms  & 10.6ms & 0.9  & 2.4  & 56.2   & 0.1    \\ \hline
  \end{tabular}

  \label{tab:segmentation evaluation}
  \vskip-4mm
\end{table}

\subsection{Recognition Results}

\begin{table} \small
  \caption{The comparison with literature baselines and reimplemented baselines. The offline AED models adopt hybrid CTC/Attention decoding algorithm. $\infty$ denotes the whole utterance.  We reproduce the TA baselines with our decoding algorithm, marked with *.  }
  \label{tab:main results}
  \centering
  \begin{tabular}{l|cc|cc}
    \toprule
    Models                & \multicolumn{2}{c|}{WER} & \multicolumn{2}{c}{latency(ms)}  \\  \cline{2-5}
                    & clean       & other      & frame  & word \\ \midrule
     \multicolumn{4}{l}{Base setting (78M)} \\ \midrule
   Contextual \cite{DBLP:conf/asru/TsunooKKW19} &  4.6   & 13.1  &  \multicolumn{2}{c}{$\infty$} \\ \hline
   TA (78M) \cite{DBLP:journals/corr/abs-2001-02674}   &  2.7  & 6.1   &\multicolumn{2}{c}{$\infty$}  \\
    TA-1             &  3.2/3.4*  & 8.2/9.5*  &    \multicolumn{2}{c}{750}         \\
    TA-2             &  2.9/3.0* & 7.8/8.5*  &  \multicolumn{2}{c}{1230}     \\
    TA-4             &  2.8  & 7.3  &  \multicolumn{2}{c}{2190}     \\   \hline
    Our method offline & 2.7  &  5.9  & \multicolumn{2}{c}{$\infty$}  \\
    ~ ~ ~ +SN-12 $\sigma$=0.5  & 3.5  & 9.4  &  317  & 63  \\
    ~ ~ ~ +SN-12 $\sigma$=0.7  &  3.3  &  8.5  & 404  & 133  \\
    ~ ~ ~ +SN-12 $\sigma$=0.9  &  2.9  &  7.4  &  619  &  338 \\
    ~ ~ ~ +SN-4 $\sigma$=0.9  &  2.8  & 7.1  & 844  & 546 \\
    ~ ~ ~ +golden      &  2.1 & 5.3  &  299  &  30  \\ \midrule
    \multicolumn{4}{l}{Large setting (140M)} \\ \midrule
    Transducer \cite{DBLP:journals/corr/abs-2002-02562} &   2.4 & 5.6   & \multicolumn{2}{c}{$\infty$}\\
    Transducer-2    &    3.0      & 7.7  & \multicolumn{2}{c}{1080}          \\
    Transducer-6    &   2.8   & 6.9      & \multicolumn{2}{c}{3240} \\ \hline
    Our method offline & 2.2  & 5.2   & \multicolumn{2}{c}{$\infty$}  \\
    ~ ~ ~ +SN-12 $\sigma$=0.9  & 2.7 & 6.4   & 639  &  352 \\ \bottomrule
  \end{tabular}

\end{table}
To measure the system latency, we define two metrics without considering the computation time, namely the frame-level latency and the word-level latency. The frame-level latency is defined as how many future frames are consumed for each frame $i$, which is same as that in  \cite{DBLP:journals/corr/abs-2001-02674}. For our model, it is computed as $(g_j - i) \times 40$ms + 30ms (CNN), where $g_{j-1} < i \leq g_j$. Word-level latency is defined as the time difference between the actual end boundary and the end boundary produced by SN. For words whose boundary is predicted correct, the word-level latency is 30ms (CNN). For words whose boundary is missed or predicted latter than the golden boundary, we compute the distance between the next boundary $g_j$ and the real boundary $g'_j$ as $(g'_j - g_j) \times 40$ms + 30ms.\footnote{The latency is computed for each word without considering subword tokens.} An example is shown in Table \ref{tab:example}.  We report the mean latency on the joint test sets. 

We compare our method with the literature baselines in Table \ref{tab:main results}.  Contextual Block  \cite{DBLP:conf/asru/TsunooKKW19} system uses a chunk-based encoder with a contextual embedding to incorporate global information and a decoder with global attention. The TA system  \cite{DBLP:journals/corr/abs-2001-02674} adopts a look ahead based encoder and a trigger-attention based decoder. In Large setting, \cite{zhang2020transformer} proposes Transformer Transducer with look-ahead based speech and label encoders.  

In the Base setting, we present model performance across different thresholds of the SN. Compared with TA baselines, our model with SN-12 and thresholds 0.9 can achieve a similar WER with their best model while reducing the latency from 2190ms to 619ms.  Model with SN-4 achieves better WER but sacrificing latency. We can see that the higher segmentation threshold (less substitution and insertion segmentation errors), the lower WER.  Tuning the threshold can give us more room for potential WER and latency tradeoffs. We also show the results when the golden segmentation is obtained. The result is surprising as it even outperforms the offline model. It is mainly because 1) the correct word boundary provides more necessary information for speech recognition, and 2) golden segmentation avoids some segmentation errors in recognition, such as ``everyday" and ``every day". In the Large setting, our model has lower WER and lower latency compared with Transformer Transducer  \cite{DBLP:journals/corr/abs-2002-02562}, which to our knowledge is the state-of-the-art results for streaming E2E ASR system on LibriSpeech benchmark.

\section{Conclusion}
In this paper, we propose a new strategy for E2E speech recognition model where a scout network detects the current word boundary and then a recognition network conducts frame-synchronous one-pass decoding by looking ahead to the predicted boundary. Our method produces a good tradeoff between WER and latency. It achieves 2.7\% and 6.4\% WER score on test-clean and test-other sets with an average of 640ms frame latency and 352ms word latency, which to our knowledge is the best publisted results for E2E streaming ASR model on LibriSpeech benchmark. 





\bibliographystyle{IEEEtran}

\bibliography{interspeech2020}


\end{document}